\documentclass[aps,preprint,showpacs,superscriptaddress,groupedaddress]{revtex4}  
\usepackage{graphicx}  
\usepackage{dcolumn}   
\usepackage{bm}        
\usepackage{amssymb}   
\usepackage{amsmath}

\usepackage{float}
\usepackage[caption = false]{subfig}
\usepackage{physics}

\begin{document}

\title{Quantum Dicke battery supercharging in the ``bound luminosity'' state}

\author{S.S. Seidov}
\affiliation{HSE University, Moscow, Russia}
\affiliation{Theoretical Physics and Quantum Technologies Department, NUST ``MISIS'', Moscow, Russia}

\author{S. I. Mukhin}
\affiliation{Theoretical Physics and Quantum Technologies Department, NUST ``MISIS'', Moscow, Russia}

\date{September 25, 2023}

\begin{abstract}

Quantum batteries, which are quantum systems to be used for storage and transformation of energy, are attracting research interest recently. A promising candidate for their investigation is the Dicke model, which describes an ensemble of two--level systems interacting with a single--mode electromagnetic wave in a resonator cavity. In order to charge the battery, a coupling between the ensemble of two--level systems and resonator cavity should be turned off at a certain moment of time. This moment of time is chosen in such a way, that the energy gets fully stored in the ensemble of two--level systems. In our previous works we have investigated a ``bound luminosity'' superradiant state of the extended Dicke model and found analytical expressions for dynamics of coherent energy transfer between superradiant condensate and the ensemble of the two--level systems. Here, using our previous results, we have derived analytically the superlinear law for the quantum battery charging power $P\sim N^{3/2}$ as function of the number $N$ of the two--level systems in the battery, and also $N$-dependence for the charging time $t_c\sim N^{-1/2}$. The $N$--exponent $3/2$ of the charging power is in quantitative correspondence with the recent result ${1.541}$ obtained numerically by other authors. The physics of the Dicke quantum battery charging is considered in detail.
\end{abstract}

\maketitle

\section{Introduction}
Quantum batteries \cite{andolina_charger-mediated_2018, le_spin-chain_2018, rossini_quantum_2020, Bhattacharjee_QB, Alicki_QB, Campaioli_QB, hu_optimal_2022, quach_superabsorption_2022} are devices which are using quantum phenomena for energy storage and its subsequent usage. Promising candidate for these devices are cavity quantum electrodynamics (CQED) systems, that are in particular described by the Dicke model \cite{Dicke, Garraway, Hepp} and a generalized extended Dicke model \cite{Rabl_2016, Rabl, Rabl_2018, Mukhin}. These systems are composed of a large number $N$ of two--level systems (TLSs) coupled to a single--mode electromagnetic resonator. The coupling enables energy exchange between the electromagnetic field and the TLSs. The collective coherent phenomena in CQED systems are responsible for various effects in the quantum batteries, such as superlinear scaling of charging power with the number $N$ of  TLSs in the battery  and inverse with $N^\alpha, \alpha>0$, scaling of quantum charging speed , etc. \cite{andolina_extractable_2019, zhang_powerful_2019, crescente_charging_2020, rosa_ultra-stable_2020, mitchison_charging_2021, crescente_enhancing_2022, seah_quantum_2021, binder_quantacell:_2015, campaioli_enhancing_2017, julia-farre_bounds_2020}.

Recently a protocol of charging of the Dicke quantum battery was proposed \cite{Gemme_QB}. It consists of turning on the coupling between the cavity and TLSs, causing the energy transfer from the electromagnetic field to the TLSs, and turning it off when the energy of the TLSs reaches its maximum. In the same work a certain approximation is made, reducing the Dicke model to the Lipkin--Meshkov--Glick (LMG) model \cite{LMG}.

In several our previous papers \cite{Mukhin_arxiv, Seidov_JETP, Seidov_Euler_top, Seidov_BL_EDM} we have described analytically semiclassical dynamics and investigated the so--called ``bound luminosity" state of the Dicke model and its generalization --- the extended Dicke model. In this state periodical beatings of the superradiant photonic condensate in the cavity happen and accordingly the energy is periodically transferred coherently from the photonic condensate to the TLSs and back. In ref. \cite{Seidov_BL_EDM} it is shown, that the ``bound luminosity" state is described by the LMG Hamiltonian, which happens to be the same as in the description of the Dicke quantum battery in \cite{Gemme_QB}. Hence, the present manuscript is devoted to application of our analytical ``bound luminosity" description to the case of Dicke quantum battery. In particular, we derive in analytical form the dependences on TLSs number $N$ of the charging time and charging power of the Dicke quantum battery and generalize these results to the case of extended Dicke model.

The paper is organized as follows. First, the extended Dicke model is introduced. Then, we briefly reproduce the results of ref. \cite{Gemme_QB} (the Dicke quantum battery) and ref. \cite{Seidov_BL_EDM} (``bound luminosity" state of the extended Dicke model) and show connection between them. Finally, we combine the two by using the analytical description of ``bound luminosity" state to calculate the energy, the charging time and the charging power of the Dicke quantum battery. 

\section{Extended Dicke model}
The Dicke model describes interaction between the members of an ensemble of $N$ two--level systems via coupling to a single--mode electromagnetic resonator cavity \cite{Dicke, Garraway, Hepp}. The extended Dicke model \cite{Rabl_2016, Rabl, Rabl_2018, Mukhin} generalizes description to the case of  an arbitrary type of direct interaction between two--level systems in the ensemble. Its Hamiltonian reads as:
\begin{equation}\label{eq:H}
H = \omega a^\dagger a + \omega_0 S_z + 2 \lambda (a^\dagger + a) S_x + (1 + \varepsilon) \frac{4 \lambda^2}{\omega} S_x^2.
\end{equation}
Here $\omega$ is the frequency of the resonator, $\omega_0$ is the energy splitting of the TLSs and $\lambda$ is the coupling constant. The parameter $\varepsilon$ defines the type of the direct interaction between TLSs. The ordinary Dicke model is obtained by setting $\varepsilon = -1$. The superspin operators $S_{x,y,z}$ are defined as:
\begin{equation}
S_{x,y,z} = \frac{1}{2}\sum_{i = 1}^N \sigma^i_{x,y,z},
\end{equation}
where $\sigma^i_{x,y,z}$ are Pauli matrices, corresponding to the $i$--th TLS in the ensemble. We also introduce the maximum total spin $S = N/2$ of the effective superspin, corresponding to the TLSs ensemble.

In the extended Dicke model with $\varepsilon \leqslant 0$ the superradiant phase transition occurs at critical coupling constant $\lambda_c$ \cite{Rabl_2016, Brandes, Seidov_JOPA}. In the superradiant phase macroscopic photonic condensate in the cavity emerges and the superspin describing TLSs' dipole moment aligns itself along the electric field $x$--axis, in the notations of Eq. (\ref{eq:H}). In particular, the averages of quantum mechanical observables become $\langle S_x \rangle = S$, $\langle q \rangle = -2 \sqrt{2} \lambda \langle S_x  \rangle /\omega$, where electromagnetic field single mode oscillator ``coordinate'' is introduced below as: $q=(a^\dagger + a)/\sqrt{2}$. The extended Dicke model in the superradiant regime is used in what follows for analytical description of the charging process of a quantum battery device.

\section{Dicke quantum battery}
The Dicke quantum battery and its charging protocol are described in ref. \cite{Gemme_QB}. The authors propose to turn on the coupling between the resonator and the TLSs ensemble at time $t = 0$ and allow the TLSs to be excited due to their interaction with the electromagnetic field. Then the coupling should be turned off at time $t_c$ (charging time), such that the TLSs ensemble reaches its maximum energy. This preserves the ensemble in the excited state not allowing its energy to be transferred back in the resonator and, hence, the battery becomes charged.

Starting with Dicke Hamiltonian, i.e. Hamiltonian (\ref{eq:H}) with $\varepsilon = -1$, the authors perform a Schrieffer--Wolf transformation. Then in dispersive regime of $\omega \gg \omega_0$ the Hamiltonian splits into free photons and an LMG model Hamiltonian:
\begin{equation}\label{eq:Heff}
H_\text{eff} \approx \omega a^\dagger a + \omega_0 S_z - \frac{4\lambda^2}{\omega} S_x^2 = \omega a^\dagger a + H_\text{LMG}.
\end{equation}

Studying the effective LMG model Hamiltonian of the spin subsystem, the authors calculate numerically and approximately analytically a number of quantities: energy stored in the battery, its average charging power and ergotropy. The energy of the battery, i.e. the energy of the TLSs ensemble, is oscillating in time.

Also from the LMG Hamiltonian the authors derive quasiclassical equations of motion by representing the spin components quasiclassically as:
\begin{equation}\label{eq:S_quasiclassic}
\begin{aligned}
&S_x = S \sin \theta \cos\varphi\\
&S_y = S \sin \theta \sin\varphi\\
&S_z = S \cos \theta
\end{aligned}
\end{equation}
and introducing canonically conjugate to $\varphi$ variable $Q = 2 S \cos \theta$, i.e. $\{\varphi, Q\} = 1$. Here the curly braces denote classical Poisson bracket, and the meaning of the last equlity is that angular momentum component along $z$-axis, $S_z$, is canonically conjugated to the rotation angle $\varphi$ around $z$-axis. The quasiclassical equations following from Hamiltonian in Eq. (\ref{eq:Heff}) then read: 
\begin{equation}\label{eq:System_Qphi}
\begin{aligned}
&\frac{d Q}{d t} = \frac{4 S^2 - Q^2}{\omega} \lambda^2 \sin(2 \varphi)\\
&\frac{d \varphi}{dt} = - \frac{\omega_0}{2} + \frac{2 \lambda^2}{\omega} Q \cos^2 \varphi.
\end{aligned}
\end{equation}
In this notation they seem complicated, but, as we show below, they can be solved exactly when approached from another perspective, and they describe the ``bound luminosity'' state of the Dicke model first introduced in \cite{Mukhin_arxiv}.

\section{``Bound luminosity" state}
A number of our previous works are devoted to description and study of a so--called ``bound luminosity" state in the ordinary and extended Dicke model \cite{Mukhin_arxiv, Seidov_JETP, Seidov_Euler_top, Seidov_BL_EDM}. In this state periodic beatings of superradiant photonic condensate in the cavity take place. The condensate is periodically absorbed and radiated by the ensemble of the TLSs coupled to the cavity. It is remarkable, that the effective Hamiltonian, describing the Dicke model in this state, is also an LMG model Hamiltonian. The quasiclassical equations of motion, describing the ``bound luminosity" state, can be solved analytically. This analytical solution provides insight in the physical processes happening in the Dicke quantum battery. In particular, the moment of time at which the coupling between the photonic condensate and TLSs should be turned off is easily calculated.

Here we briefly reiterate calculations from ref. \cite{Seidov_BL_EDM}, where extended Dicke model Hamiltonian is transformed into the LMG model Hamiltonian and  subsequently the quasiclassical equations of motion are solved. First we introduce the electromagnetic field ``coordinate'' and ``momentum'' operators $q = (a + a^\dagger)/\sqrt{2}$ and $p = i (a^\dagger - a)/\sqrt{2}$. Then the Hamiltonian (\ref{eq:H}) can be written as:
\begin{equation}
H = \frac{\omega}{2}(q^2 + p^2) + \omega_0 S_z + 2 \sqrt{2} \lambda q S_x + (1 + \varepsilon) \frac{4 \lambda^2}{\omega} S_x^2.
\label{ini}
\end{equation}
This Hamiltonian leads to quasiclassical equations of motion of superradiant condensate coupled to TLSs:
\begin{equation}\label{eq:System}
\begin{aligned}
&\dot S_x = - \omega_0 S_y\\
&\dot S_y = \omega_0 S_x - 2 \sqrt{2} \lambda q S_z - (1 + \varepsilon) \frac{8 \lambda^2}{\omega} S_x S_z\\
&\dot S_z = 2 \sqrt{2} \lambda q S_y + (1 + \varepsilon) \frac{8\lambda^2}{\omega} S_x S_y\\
&\dot q = \omega p\\
&\dot p = -\omega q - 2\sqrt{2} \lambda S_x.
\end{aligned}
\end{equation}
The system has the following fixed points in the phase space $(q, p, S_x, S_y, S_z)$:
\begin{equation}\label{eq:Fixed_pts}
\begin{aligned}
&\mathbf{x}_0^\text{pole} = \begin{pmatrix}
0, &0, &0, &0, &\pm S
\end{pmatrix}\\
&\mathbf{x}_\pm = \begin{pmatrix}
\pm \dfrac{2 \sqrt 2 \lambda}{\omega} \sqrt{S^2 - S_z^2}, &0, &\mp \sqrt{S^2 - S_z^2}, &0, &\dfrac{\omega \omega_0}{8 \varepsilon \lambda} 
\end{pmatrix}.
\end{aligned}
\end{equation} 
The fixed point $\mathbf{x}_{0}^\text{pole}$ corresponds to the normal phase. The fixed points $\mathbf{x}_\pm$ appear at $\lambda > \lambda_c = \sqrt{\omega \omega_0/(8 S |\varepsilon|)}$. The latter condition guarantees the real value of the square roots  in the expressions above and corresponds to the superradiant phase of the extended Dicke model. These fixed points correspond to two symmetric superradiant states with nonzero values of $q = -{2 \sqrt{2} \lambda}/{\omega} S_x $ and $S_x=\pm S$. Accordingly, $\lambda_c$ is the critical coupling constant at which the superradiant phase transition occurs. The origin of the latter fixed points becomes obvious when one completes a full square \cite{Seidov_JOPA} in Hamiltonian (\ref{ini}):
\begin{equation}
H = \frac{\omega}{2}p^2 +\frac{\omega}{2} \left(q +\frac{2 \sqrt{2} \lambda}{\omega} S_x \right)^2+ \omega_0 S_z +  \frac{4 \varepsilon \lambda^2}{\omega} S_x^2.
\label{fini}
\end{equation}
Hence, we now consider the case of superradiant photonic coordinate $q$, that follows adiabatically \textbf{(i.e. $\dot q \approx 0$, $\dot p \approx 0$)} the $S_x$ component of the superspin at the bottom of the  parabolic potential well of photonic oscillator, expressed by the second term in Eq. (\ref{fini}): $q \approx - 2 \sqrt{2} \lambda S_x / \omega$ and $p \approx 0$. Substituting these approximations back into equations (\ref{eq:System}) we obtain a system of equations of motion for the spin variables:
\begin{equation}\label{eq:System_reduced}
\dot S_x = - \omega_0 S_y 
\end{equation}
\begin{equation}\label{eq:System_reduced1}
\dot S_y = \omega_0 S_x - \frac{8 \varepsilon \lambda^2}{\omega} S_x S_z 
\end{equation}
\begin{equation}\label{eq:System_reduced2}
\dot S_z = \frac{8 \varepsilon \lambda^2}{\omega} S_x S_y.
\end{equation} 
Remarkably, this system is generated by the Hamiltonian:
\begin{equation}\label{eq:HLMG}
H_\text{LMG} = \omega_0 S_z + \frac{4 \varepsilon \lambda^2}{\omega} S_x^2\,,
\end{equation}
which is precisely the LMG model Hamiltonian in equation (\ref{eq:Heff}) when one takes $\varepsilon = -1$. The system of equations (\ref{eq:System_reduced})--(\ref{eq:System_reduced2}) with $\varepsilon = -1$ is also equivalent to the system (\ref{eq:System_Qphi}) when the quasiclassical representation of spin (\ref{eq:S_quasiclassic}) is taken into account.

\begin{figure}[h!!]
\center\includegraphics[width=0.75\textwidth]{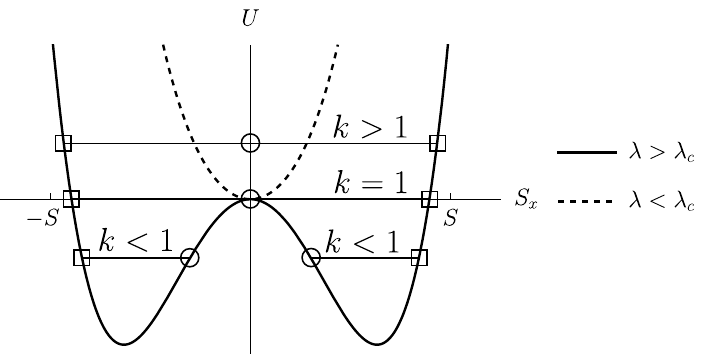}
\caption{The effective double--well potential (solid line) $U(S_x)$ in Eq. (\ref{eq:Jac}) corresponding to ``bound luminosity" superradiant state. The normal state potential is the dashed line. Horizontal lines mark the total ``energy" $C$ being above the potential's maximum ($k > 1$, $S_x \sim \operatorname{cn}(\Omega t, k)$) and below the maximum ($k < 1$, $S_x \sim \operatorname{dn}(\Omega t, k)$). The $S_x$ axis itself corresponds to the ``energy" $C=0$ at the potential's maximum ($k = 1$, $S_x \sim \operatorname{sech}(\Omega t, k)$). The points on the potential energy curve that correspond to the quantum battery lowest and highest charging energies $\sim\omega_0S_z$ are marked with empty squares and circles accordingly. Note, that although it does not follow from the plot, the coordinate $S_x$ can not exceed $S$ due to total spin conservation law.}
\label{fig:U}
\end{figure}

Now, substituting $S_y$ from Eq. (\ref{eq:System_reduced}) into (\ref{eq:System_reduced2}) one obtains an expression which is a full differential, that after one integration over time reads:
\begin{equation}\label{eq:HLMG0}
\omega_0 S_z + \frac{4 \varepsilon \lambda^2}{\omega} S_x^2=E.
\end{equation}
The lhs of Eq. (\ref{eq:HLMG0}) actually coincides with Hamiltonian in Eq. (\ref{eq:HLMG}), modulo $c$-numbers $S_{z,x}$ instead of operators are used.
Now, differentiating with respect to time both sides of Eq. (\ref{eq:System_reduced}) and then eliminating $S_y$, $S_z$ variables from Eq. (\ref{eq:System_reduced1}), using Eqs. (\ref{eq:System_reduced}) and (\ref{eq:HLMG0}) one finally finds dynamics equation for the $S_x$ component coinciding with equation for Jacobi elliptic functions \cite{Whittaker}: 
\begin{equation}\label{eq:Jac}
{\dot S_x}^2=C + \left(\frac{8\varepsilon\lambda^2}{\omega}E-\omega_0^2 \right)S_x^2-\frac{16\varepsilon^2\lambda^4}{\omega^2}S_x^4\equiv C-U(S_x).
\end{equation} 
Thus, the rhs of (\ref{eq:Jac}) contains an effective double--well potential energy $U(S_x)$ and $C$ is an integration constant, i.e. the ``energy", see Fig. \ref{fig:U}.
The Eq. (\ref{eq:Jac}) is exactly solvable \cite{Whittaker}:
\begin{equation}\label{eq:sol_Sx}
S_x(t) = \pm \frac{\omega \Omega}{4  \lambda^2 |\varepsilon|} \operatorname{dn}(\Omega t, k)\,,
\end{equation}
where $\operatorname{dn}(\Omega t, k)$ is the Jacobi elliptic function, and integration constant is written in the form:
\begin{equation}\label{eq:C}
C=(k^2-1)\frac{\omega^2\Omega^4}{16\varepsilon^2\lambda^4}.
\end{equation}
The frequency $\Omega$ is found from Eq. (\ref{eq:Omega}) below, that follows from the total superspin $S$ conservation law. Then, for the Jacobi parameter $k$ one has consistency condition following from the coefficient in front of $S_x^2$ in Eq. (\ref{eq:Jac}):
\begin{equation}\label{eq:sol_Sxk}
(2-k^2)\Omega^2 = \frac{8 \varepsilon \lambda^2 E}{\omega}-\omega_0^2.
\end{equation} 
The rest of the spin components can be found using relations (\ref{eq:System_reduced}) and (\ref{eq:HLMG0}).
An equation for $\Omega$ follows from the total spin conservation law, that remarkably turns into time--independent relation after substitution of the $S_{x,y,z}(t)$ solutions found above:
\begin{equation}\label{eq:S2}
S_x^2(t) + S_y^2(t) + S_z^2(t) = \frac{1}{16 \varepsilon^2 \lambda^4} \left[\omega^2 \Omega^2 + \frac{(4 \varepsilon \lambda^2 E - \omega \Omega^2)^2}{\omega_0^2} \right] = S^2.
\end{equation}
Substituting into Eq. (\ref{eq:S2}) constant $E$ from Eq. (\ref{eq:sol_Sxk}) one finds equation that defines the frequency $\Omega$:
\begin{equation}\label{eq:Omega}
\Omega^4 + 2 \omega_0^2 \Omega^2 \frac{2-k^2}{k^4} -64\frac{\varepsilon^2\lambda^4\omega_0^2 S^2}{k^4\omega^2}+\frac{\omega_0^4}{k^4}=0,
\end{equation}
that leads to the following asymptoticsolution in the limit $N\gg 1$:
\begin{equation}\label{eq:Omega1}
\Omega \Big|_{S \rightarrow \infty} = \frac{2 \sqrt 2 \lambda}{k} \sqrt\frac{|\varepsilon| \omega_0 S}{\omega}.
\end{equation}
Hence, the frequency $\Omega$ grows $\sim \sqrt{|\varepsilon| \omega_0 S}$ at $N\gg 1$. The only variable parameter in the problem is now Jacobi function modulus $k$ entering periodic in time $t$ solution for $S_x(t)$ in Eq. (\ref{eq:sol_Sx}). The latter, depending on value of $k$, describes either oscillations around one of the fixed points $\mathbf{x}_\pm$, corresponding to two symmetric superradiant phases, or meandering between them. This dynamic state we call the ``bound luminosity'' state. It is remarkable, that biquadratic equation Eq. (\ref{eq:Omega}) has real solutions $\Omega(k)$ only at $\lambda > \lambda_c = \sqrt{\omega \omega_0/(8 S |\varepsilon|)}$, which simultaneously serves as the condition of  transition into the superradiant phase found above.

\section{Superlinear charging power in the ``bound luminosity" state}
In the previous section we have shown that the ``bound luminosity" regime of the Dicke model is described by the same LMG model Hamiltonian as the Dicke quantum battery in \cite{Gemme_QB}. Here our description is generalized to the extended Dicke model, i.e. instead of fixing $\varepsilon=-1$ as in the case considered before \cite{Gemme_QB}, we consider an arbitrary coefficient $\varepsilon<0$. Charging power of the quantum Dicke battery is calculated below analytically and it follows  superlinear dependence $\sim N^{3/2}$  on the number $N\gg 1$ of TLSs  in accord with $N^{1.541}$ dependence found numerically in \cite{Gemme_QB}. The charging power is defined as $P=dE_B(t)/dt$, where the energy of the battery $E_B(t)$ at time $t$ is expressed via the energy of the two--level systems in the cavity:
\begin{equation}\label{eq:Ez}
\begin{aligned}
E_B(t) = E_Z(t) - E_Z(0);\;\; E_Z(t) =\omega_0 S_z(t)= E + \frac{\omega \Omega^2}{4 |\varepsilon| \lambda^2}\operatorname{dn}^2(\Omega t, k).
\end{aligned}
\end{equation} 
Here we have expressed $S_z(t)$ using equations (\ref{eq:HLMG0}) and (\ref{eq:sol_Sx}). Simultaneously, the dipole energy of spin--photon interaction equals $E_\text{dip}(t) \sim \lambda q(t) S_x(t)$ in the rhs of the Dicke Hamiltonian (\ref{ini}). In the ``bound luminosity" state beatings of the photonic condensate intensity occur due to periodic transfer of energy between TLSs and photonic condensate subsystems via $E_\text{dip}(t)$ term \cite{Mukhin_arxiv, Seidov_BL_EDM}. Solution (\ref{eq:sol_Sx}) allows to find the charging time $t_c$, as well as charging power $P$. 

\begin{figure}[h!!]
\center\includegraphics[width=0.72\textwidth]{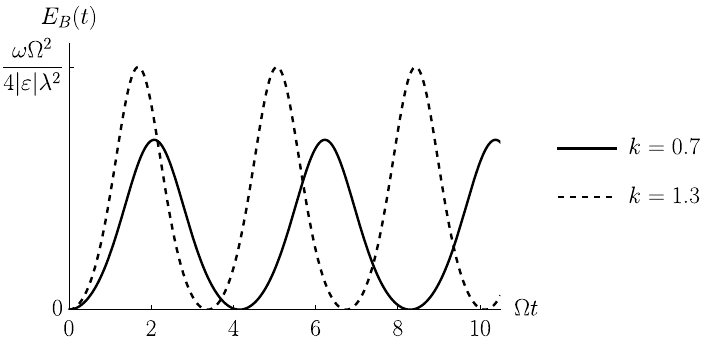}
\caption{The energy of the Dicke battery as function of time for different values of parameter $k$, $\varepsilon < 0$.}
\label{fig:EB}
\end{figure}

One should choose $\varepsilon < 0$ in order to assure the existence of the superradiant condensate, then expression (\ref{eq:Ez}) is maximal at $t = 0$. Hence, in order to describe battery charging process we shift time origin $t=0$ to the minimum of the expression in Eq. (\ref{eq:Ez}), when the TLSs are in their ground state:
\begin{equation}
E_Z(t) =  E - \frac{\omega \Omega^2}{4 \varepsilon \lambda^2}\operatorname{dn}^2 \big(\Omega t + K(k), k \big),
\end{equation} 
where $K(k)$ is the complete elliptic integral of the first kind \cite{Whittaker}. Then, for the energy of the battery at time $t$ one obtains:
\begin{equation}\label{Et}
E_B(t) = E_Z(t) - E_Z(0) = \frac{\omega \Omega^2}{4 |\varepsilon| \lambda^2} \left[ \operatorname{dn}^2\big(\Omega t +K(k), k \big) - 1 + k^2\right].
\end{equation}
This function is plotted in Fig. \ref{fig:EB}. At $t = 0$ Jacobi function $\operatorname{dn}^2(K(k), k) = 1 - k^2$ and $E_B(0) = 0$, the battery is discharged. Then, $E_B(t)$ is maximal when $\operatorname{dn}^2 \big(\Omega t + K(k), k \big)$ reaches its maximum value of $1$, which gives the condition for the charging time:
\begin{equation}
\operatorname{dn}^2 \big(\Omega t_c + K(k), k \big) =\operatorname{dn}^2 \big(2K(k), k \big) = 1\; \Rightarrow t_c = \frac{K(k)}{\Omega}.
\end{equation}
These derivations are valid for $k < 1$. In the case of $k > 1$ we are using the relation \cite{Whittaker}:
\begin{equation}\label{eq:cn}
\operatorname{dn}(\Omega t, k) = \operatorname{cn}\left(\Omega{k} t, \frac{1}{k} \right)
\end{equation}
and obtain instead of (\ref{Et}):
\begin{equation}
\begin{aligned}
&E_B(t)\Big|_{k > 1} = \frac{\omega \Omega^2}{4 |\varepsilon| \lambda^2}\operatorname{cn}^2 \left(\Omega k t+K(1/ k), \frac{1}{k} \right)\\
& t_c\Big|_{k > 1} = \frac{K(1/ k)}{\Omega k}.
\end{aligned}
\end{equation}
For $k = 1$ the Jacobi functions turn into the $\operatorname{sech}(\Omega t)$ function, i.e. $\operatorname{dn}(\Omega t, 1) = \operatorname{cn}(\Omega t, 1) = \operatorname{sech}(\Omega t)$. In this case $t_c \approx \Omega^{-1}$ is defined as the characteristic time during which function $\operatorname{sech} (\Omega t)$ changes from $\ll 1$ to $1$.

Finally, the maximal energies of the battery $E_B(t_c)$ and the charging times $t_c$ are:
\begin{equation}
\begin{aligned}
E_B(t_c) &= k^2 \dfrac{\omega \Omega^2}{4 |\varepsilon| \lambda^2}&  t_c &= \dfrac{K(k)}{\Omega}& k &< 1\\
E_B(t_c) &= \dfrac{\omega \Omega^2}{4 |\varepsilon| \lambda^2}&  t_c &= \dfrac{K(1/k)}{\Omega k}&  k &> 1.
\end{aligned}
\end{equation}
As one can see from above equations, the battery full charge does not contain $k$-dependent prefactor for $k > 1$. By differentiating $E_B(t)$ in eq. (\ref{Et}) with respect to time we find for the charging power
\begin{equation}
P(t) = \frac{d E_B(t)}{dt} = \frac{\omega \Omega^3 k}{2 |\varepsilon| \lambda^2} f(\Omega t+K(k)),
\end{equation}
where $f(t) \sim \operatorname{cn}(t) \operatorname{sn}(t) \operatorname{dn}(t)$ is a function, which does not exceed unity. From (\ref{eq:Omega1}) one finds $\Omega \sim \sqrt S$ and thus:
\begin{equation}\label{PQB1}
P \sim S^{3/2}.
\end{equation}
Given that $S = N/2$, where $N$ is the number of TLS in the system, the obtained result is remarkable, as it demonstrates that in the ``bound luminosity" state of the superradiant photonic condensate and TLS ensemble in a QED cavity the system described by the extended Dicke model possesses supercharging property, i.e. the charging power is superlinear in the TLS number $N$. Simultaneously, the charging time $t_c$ decreases with the TLS number $N=2S$ as $N^{-1/2}$:
\begin{equation}\label{PQB12}
t_c \sim \Omega^{-1} \sim S^{-1/2} \sim N^{-1/2}.
\end{equation}
It is also remarkable, that analytical result in Eq. (\ref{PQB1}) for a general extended Dicke model ($-1 \leqslant \varepsilon < 0$) is in a quantitative correspondence with numerical results reported in \cite{Gemme_QB} for the Dicke model $\varepsilon=-1$, see Fig. 4b therein, where $P\sim N^{1.541}$ was found.
Finally, the distinction between the cases $k^2\lesseqgtr 1$ seen in formulas (\ref{Et}) and (\ref{eq:cn}) arises due to different positions of the ``energy" $C$, given by (\ref{eq:C}), with respect to the barrier top of the effective double--well potential energy $U(S_x)$ of the Jacobian dynamic equation (\ref{eq:Jac}).
Namely, the cases $k^2<1$, $k^2>1$ and $k^2=1$ correspond respectively to the ``energy" $C$ below, above and at the top of the potential energy maximum at $S_x=0$, that separates the two symmetrical wells of $U(S_x)$ along the $S_x$ axis, see Fig. \ref{fig:U}.  

\section{Conclusions}
In the present paper we have derived in analytical form dynamic characteristics of the Dicke quantum battery using our previous results that describe the ``bound luminosity" state of the extended Dicke model. By solving exactly the qusiclassical equations of motion for the TLS and superradiant photonic condensate we have derived analytical expressions for the energy $E_B$, charging time $t_c$ and charging power $P$ of the Dicke battery. Obtained here analytical expressions indicate that charging time of the Dicke quantum battery in the ``bound luminosity" state decreases with TLS number $N$ as $t_c\sim N^{-1/2}$. Simultaneously, found analytically (super)charging power dependence $P\sim N^{3/2}$ is in quantitative correspondence with the result $P\sim N^{1.541}$ obtained numerically for the superradiant state of the Dicke model in \cite{Gemme_QB}. 

\section{Acknowledgements} 
S.I.M. acknowledges support by the Federal Academic Leadership Program Priority 2030 (NUST MISIS Grant No. K2-2022-025). S.S.S. acknowledges support from the Basic Research Program of HSE University project TZ-15 ``Nanoelectronics of low dimensional systems''.

\bibliographystyle{ieeetr}
\bibliography{biblio}

\end{document}